\documentclass[%
reprint,
superscriptaddress,
nobibnotes,
nolongbibliography,
amsmath,amssymb,
aps,
prb,
]{revtex4-2}

\usepackage{graphicx}
\usepackage{dcolumn}
\usepackage{bm}
\usepackage{siunitx}
\usepackage{placeins}
\usepackage{hyperref}

\newcommand{\figref}[1]{Figure~\ref{#1}}

\begin{document}
\title{Long-lived Photoluminescence of Photostable One-dimensional Picoperovskites}
\author{Maximilian Tomoscheit}
\affiliation{Institute of Physics, University of Rostock, 18059 Rostock, Germany}
\affiliation{Institute of Solid State Physics, Friedrich Schiller University Jena, 07743 Jena, Germany}
\author{Julian Schröer}
\affiliation{Institute of Physics, University of Rostock, 18059 Rostock, Germany}
\author{Jaskaran Singh Virdee}
\affiliation{Institute of Physics, University of Rostock, 18059 Rostock, Germany}
\author{Rico Schwartz}
\affiliation{Institute of Physics, University of Rostock, 18059 Rostock, Germany}
\author{Christopher E. Patrick}
\affiliation{Department of Materials, University of Oxford,  Oxford OX1 3PH, United Kingdom}
\author{Reza J. Kashtiban}
\affiliation{ Department of Physics,  University of Warwick, Coventry, United Kingdom}
\author{Tobias Korn}
\affiliation{Institute of Physics, University of Rostock, 18059 Rostock, Germany}

\begin{abstract}
We study one-dimensional metal halide perovskite crystals encapsulated in single-wall nanotubes, so-called picoperovskites, using optical spectroscopy. Polarized micro-photoluminescence (PL) reveals bright emission from aligned bundles of picoperovskites with clear linear polarization along the bundle axis. This emission is red-shifted with respect to bulk perovskite samples using the same constituents. Temperature-dependent, time-resolved micro-PL shows extraordinarily long PL lifetimes of the picoperovskites at low temperatures, reaching several hundred nanoseconds and exceeding those of bulk perovskites by two orders of magnitude.
\end{abstract}
\maketitle

\section{Introduction}
Perovskite materials, defined by the sum formula ABX$_3$, have sparked significant research interest in recent years, particularly due to their excellent properties as solar cell materials~\cite{Snaith-Solar}. However, they also show other exciting properties, such as huge Rashba splitting~\cite{Park2019} due to strong spin-orbit coupling, large nonlinear susceptibility values~ \cite{Blancon2020} and near-unity quantum yield~\cite{Dutt2022}. Perovskite materials can be easily synthesized in various forms with diverse chemical compositions, including inorganic, organic, and hybrid perovskites. The so-called metal halide perovskites (MHPs) have been extensively studied and have been shown to have tunable emission wavelengths via chemical composition, dimensionality and layer number~\cite{Ha2017}. Overall, the properties of perovskites are strongly correlated with their structure, e.g., bond angles and lengths between the metal and halide atoms. These values can change with temperature, leading to temperature-dependent properties and drastic phase changes. Layered two-dimensional (2D) perovskites can be synthesized or exfoliated down to monolayer thickness.This reduced dimensionality leads to significant confinement effects, like increasing the optical bandgap and the exciton binding energies~\cite{Blancon2018}. All of this makes them interesting candidates for fundamental research and device applications, especially for solar cells and optoelectronic devices. A major challenge for device application are the pronounced degradation processes of perovskites, which have been intensely investigated in recent years; see, e.g., Refs.~\cite{Bisquert-degradation,Wei-degradation} and references therein. Degradation may be driven by the presence of oxygen and moisture. Additionally, under light exposure, photodegradation takes place, which is especially problematic for photovoltaic applications. Photodegradation also hinders basic research of the perovskite materials themselves, especially when structures with reduced dimensionality and correspondingly greater surface-to-volume ratio, such as layered perovskites, are to be investigated by optical spectroscopy. 
Several strategies were employed to reduce  the impact of photodegradation, such as measurements and storage in inert atmosphere, performing measurements at cryogenic temperatures, or encapsulation of 2D perovskites between few-layer flakes of hexagonal boron nitride (hBN)~\cite{Fang-encapsulation,Seitz-encapsulation}.  None of these measures can stop degradation indefinitely, making long-term storage of reduced-dimensionality perovskite materials and their application in devices difficult. Temperature-dependent optical experiments are also challenging, as samples typically experience severe degradation after a single cooling cycle, even though the experiments are carried out in a vacuum or inert atmosphere.  

Here, we study one-dimensional (1D) picoperovskites encapsulated inside single-walled carbon nanotubes (SWCNTs) by various optical spectroscopy techniques. In these structures, the SWCNTs act, both, as a scaffold that yields the smallest conceivable 1D perovskite nanowire, and as a near-perfect encapsulation barrier, providing protection against degradation mechanisms. We find bright photoluminescence (PL) emission from bundles of aligned picoperovskites with a clear linear polarization along the bundle axis. This emission is red-shifted with respect to bulk perovskites using the same constituents. Temperature-dependent, time-resolved micro-PL shows extraordinarily long PL lifetimes of the picoperovskites at low temperatures, reaching several hundred nanoseconds and exceeding those of bulk perovskites by two orders of magnitude. The picoperovskites display remarkable stability regarding long-term storage, cooling cycles and optical excitation. 

\begin{figure}
    \centering
\includegraphics[width=\linewidth]{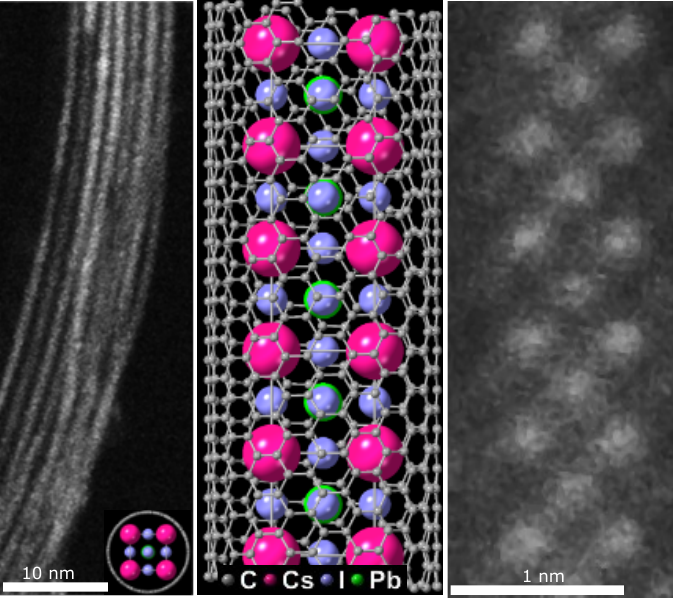}
    \caption{Annular dark-field scanning transmission electron microscopy (ADF-STEM) micrograph of CsPbI nanowires encapsulated within a bundle of SWCNTs with inset depicting end-on view of a nanowire structure model (left). A representative structure model of hybrid CsPbI Picoperovskite/ SWCNT (middle) and ADF-STEM view of a 0.8 nm in diameter nanowire within a 1.4 nm SWCNT. The scale bars from left to right represent 10 nm and 1 nm respectively.}
    \label{fig:per-uc}
\end{figure}

\section{Results and Discussion (PL characterization)}
We study picoperovskite samples which consist of 1D CsPbI$_3$ perovskite nanowires enclosed by Single Wall Carbon Nanotubes (SWCNT). Due to the small scaffold provided by the SWCNTs, the nanowires have a diameter below one nanometer, hence the combined SWCNT-perovskite nanostructures will be called picoperovskites, in line with previous work~\cite{Kashtiban2023}. Growth of the perovskites within the SWCNT is realized by mixing finely ground perovskite powder with oxidized SWCNTs. The mixture is then sealed in an ampoule under vacuum,  slowly heated to about 500° C, kept at elevated temperature for about 12 hours and then slowly cooled to room temperature. This process yields a powder containing picoperovskites. Further details of the preparation are reported elsewhere~\cite{Kashtiban2023}. High-resolution transmission electron microscopy images of picoperovskites, together with a structure model, are depicted in \figref{fig:per-uc}.

 Before sample fabrication, the source material was stored under ambient conditions with no extra precautions. To prepare samples for optical spectroscopy, small amounts of the picoperovskite powder were mixed into isopropanol, and the mixture was sonicated for several minutes to disperse entangled bundles. The sonicated solution is then quickly dropped onto a Si/SiO$_2$ wafer and spin-coated to ensure a low-density coverage of individual picoperovskites and bundles. Reference samples were prepared in a similar fashion, but using bulk CsPbI perovskite instead of picoperovskite powder to create mesoscopic agglomerates of bulk perovskite on a Si/SiO$_2$ wafer. To test the stability, the samples were exposed to a laser for several minutes while continously taking PL spectra. For a moderate laser power no systemic reduction of the PL signal occured, the small changes observed are attributed to power drift of the laser and heat-induced spatial drift of the sample position. This lack of degradation is in contrast to the results found for a 3D sample. In oder to induce comparable degradation in the picoperovskite sample, an increase of laser power by more than a factor of 3 is required (see SI Figure 3). In between measurements the samples were stored under ambient conditions and were reused for multiple experiments and cooling cycles.
 
\begin{figure}[h]
    \centering
    \includegraphics[width=\linewidth]{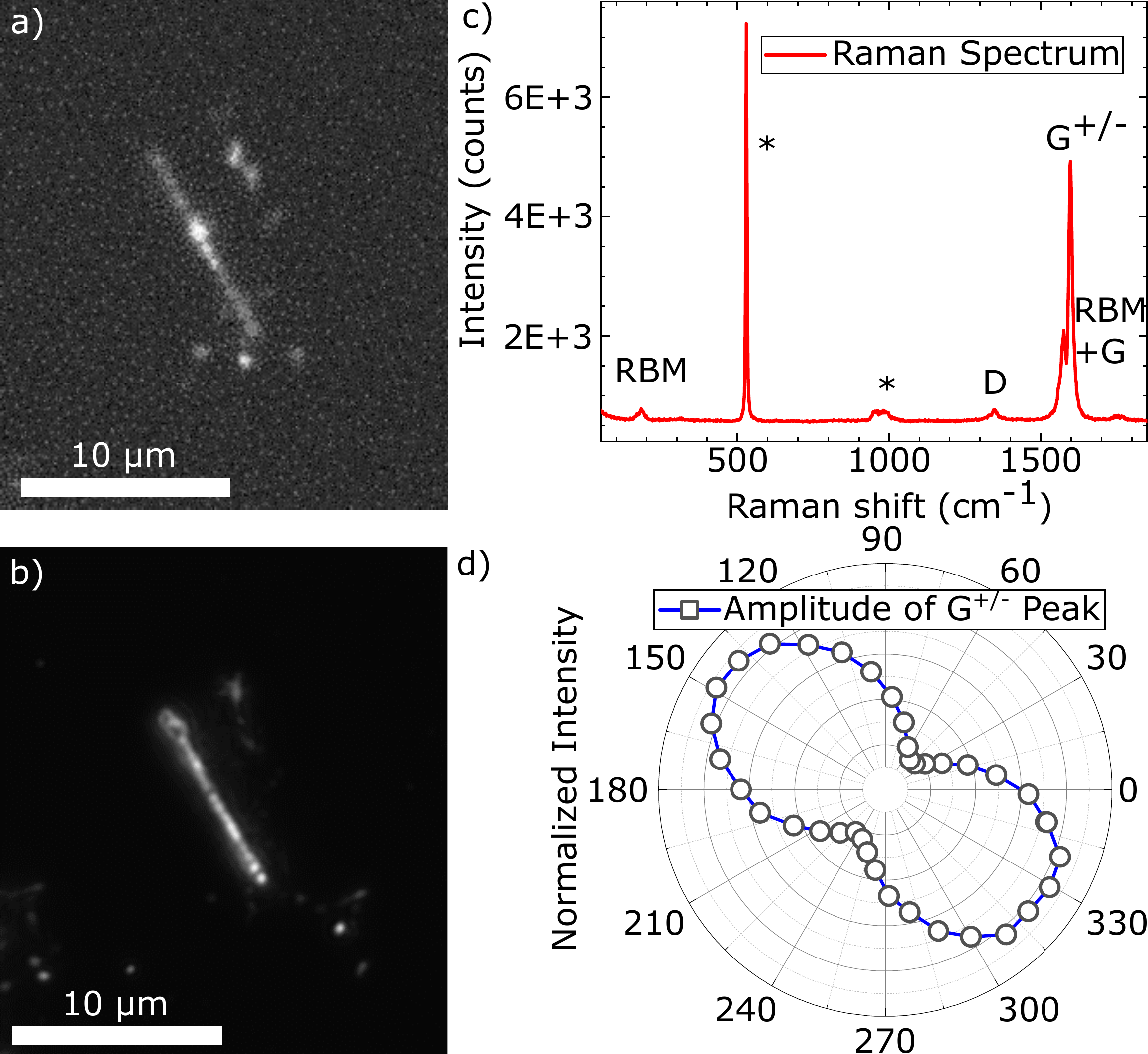}
    \caption{a) Fluoresence of a picoperovskite bundle. The sample is illuminated by a spectrally filtered (380-420~nm) LED source, the fluorescence is filtered by a 450~nm longpass.  b) Dark field image of the same area as shown in a). c) Raman spectrum of a picoperovskite bundle. d) Polarization-resolved intensity of G$^+$ peak measured with fixed excitation polarization.}
    \label{fig:fluo-raman}
\end{figure}

 For basic characterization, fluorescence (FL) and dark-field (DF) images obtained with a Nikon microscope under 50x magnification were used. For FL imaging, a Nikon V-2A filter block is utilized. It transmits blue light (380-420~nm) from a high-power light-emitting diode source onto the sample. The light reflected and emitted from the sample is filtered via a long pass (onset wavelength 450~nm) filter and detected using a Peltier-cooled charge-coupled device (CCD) camera. \figref{fig:fluo-raman} a) and b) show the FL and DF images taken on one sample examined in this study. Both images show a wire-like structure with a length of about 9 microns and an apparent width of about 700 nm, limited by the spatial resolution of the optical microscope. In the DF image, this structure is the only signal source, whereas in the FL image, the structure is surrounded by other fluorescing sources, possibly organic residues.
An integration time of \qty{10}{\second} was used to record the FL signal. In comparison between a) and b), the FL signal of the structure is uniformly distributed, except for a local maximum roughly situated in the middle, which we will see later in a different sample is not due to local enhancement but likely originates from organic residues overlapping the structure. The DF image in b) on the other hand is quite inhomogeneous, possibly due to a varying thickness or roughness of the structure. 

To identify the composition of the structure Raman measurements were carried out. The sample was excited by a \qty{532}{\nano\meter} laser with a fixed power of \qty{1}{\milli\watt}. For polarization-resolved measurements, a linear polarizer and a half-wave plate were placed before the spectrometer. The  Raman spectrum in \figref{fig:fluo-raman} c), taken at an arbitrary polarization angle, shows the typical Raman signature at low and intermediate Raman shifts expected for a SWCNT\cite{Dresselhaus2002}. The small peak at  low energy (\qty{128,9}{\per\centi\meter}) is the radial breathing mode (RBM) from which a diameter of \qty{1.36}{\nm} can be derived\cite{Maultzsch-RBM}. This is in good agreement with the CNT diameters used in the synthesis of the picoperovskites~\cite{Kashtiban2023}, which are in the range of \qty{1}{}-\qty{2}{\nano\meter}. From this range of CNTs, only a subset that has interband transitions resonant with the fixed laser wavelength yields sufficient Raman intensity to be easily detected by our Raman setup, leading to a single RBM peak. Furthermore, in the mid-energy range, four more peaks can be identified, the defect peak D, the in-plane bond stretching modes G$^{+/-}$, with the plus and minus components emerging due to strain effects related to the curvature of the SWCNTs, and the combined RBM+G mode. The high ratio of the G$^+$ peak to the D peak ($G/D\sim 36.5$) indicates a low defect density in the SWCNT. 

For the polarization-dependent Raman signal amplitude pattern shown in d), the G$^{+}$ peak obtained at different detection polarization angles was fitted by a single Gaussian. The resulting pattern is that of a two-fold symmetric system, as previously reported for individual SWCNTs~\cite{Duesberg-PolRamanCNT}.
This arises from the vastly different polarizability of CNTs for external fields applied parallel and perpendicular to the tube axis, which in turn modulates the absorption of light depending on its polarization orientation with respect to the tube axis by a factor of about 20~\cite{AJIKI1994349}.
For a single SWCNT, we would thus expect the Raman signature to almost vanish perpendicular to the long axis. However, in the measured polarization pattern the signal only gets reduced by \qty{83}{\percent}, less than expected for a single isolated SWCNT. Thus, we infer that our structure consists of multiple SWCNTs forming an aligned bundle with a small divergence of the individual tube axes, in good agreement with the inhomogeneity observed in the DF image.
Since SWCNTs with diameters used in our synthesis have band gaps corresponding to near-infrared PL~\cite{PL-CNTs}, the fluorescence in the visible range observed in our structure must stem from the perovskite core. Its optical properties will be studied in the following.

\begin{figure}[h]
    \centering
    \includegraphics[width=\linewidth]{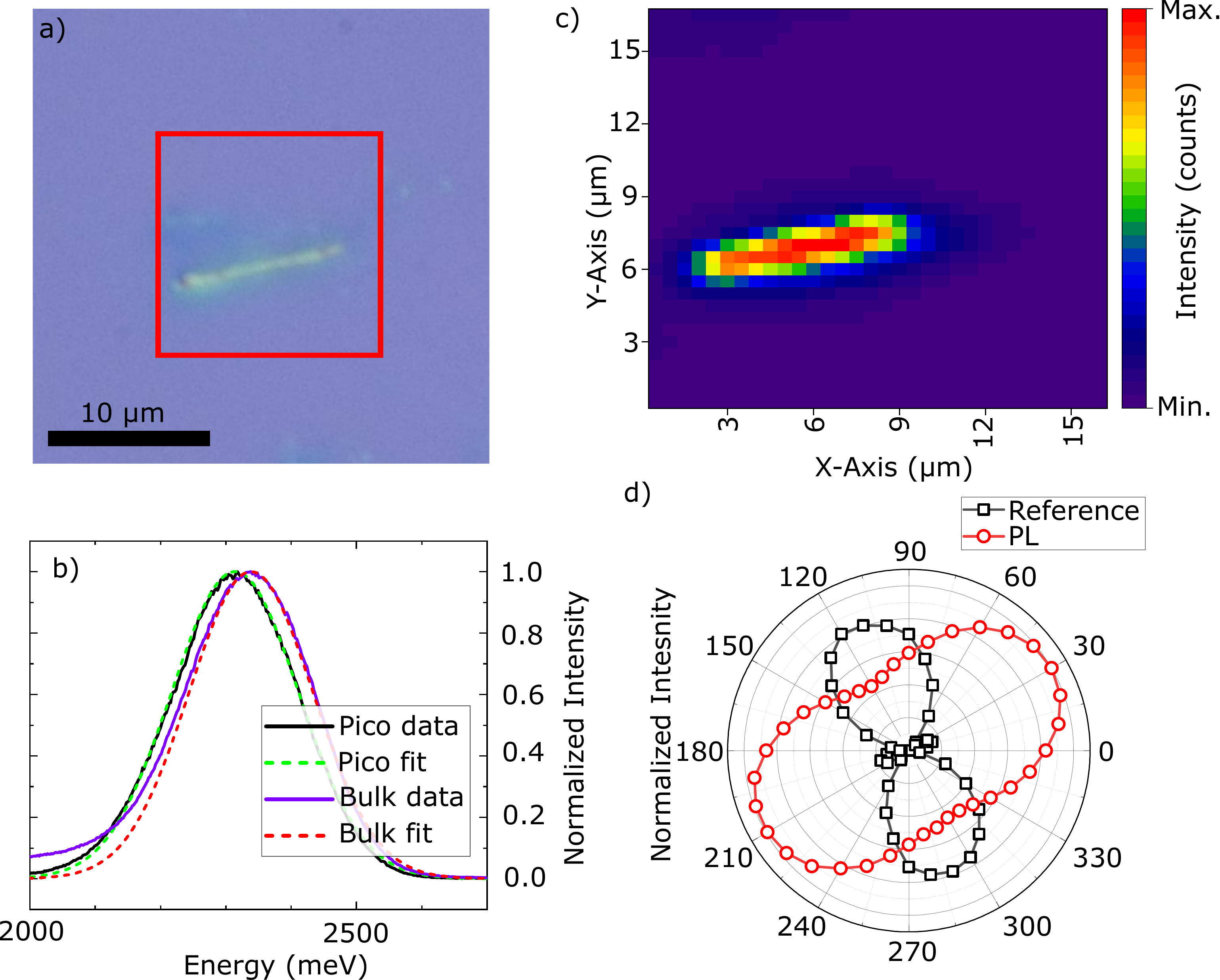}
    \caption{a) Optical image of a picoperovskite bundle. b) PL spectra (solid line) and Gaussian fits (dashed lines) obtained at T=4 K of the picoperovskite bundle shown in a) and bulk sample for comparison. c) False color map of PL intensity extracted from a scan of the area marked by the red box in a). d) Polarization-resolved intensity of backscattered excitation laser light (black squares) and picoperovskite PL emission.}
    \label{fig:opto-pl}
\end{figure}

An optical microscope image recorded with epi-illumination of another picoperovskite sample can be seen in Fig.\ref{fig:opto-pl} a).  The slightly irregular shape is likely due to the bundle consisting of mostly aligned picoperovskites  with varying lengths and diameters. To further study the fluorescence a PL setup with a diode laser emitting at \qty{405}{\nano\meter} was used. An automated stage was used to obtain PL maps and a combination of polarizer and half wave plate in front of the spectrometer, as in the Raman setup, to obtain polarization-resolved spectra. \figref{fig:opto-pl} b) shows two individual spectra measured at \qty{4}{\kelvin} and an arbitrary polarization angle with corresponding Gaussian fits in dotted lines. The purple line is the PL spectrum obtained from a bulk reference sample under the same experimental conditions. In comparison the spectrum obtained from the picoperovskite sample, shown in black, is slightly red shifted to lower energies by \qty{36,5}{\milli\electronvolt}, centered at \qty{2.313}{\electronvolt}.

Given the drastic change of dimensionality from a bulk-like to a 1D perovskite structure, this small energy shift is surprising. However, we have to consider that there are multiple effects that influence the optical band gap upon this change that work in opposite directions~\cite{Wu_2017}. First,  size quantization increases the interband transition energy, as electron and hole energy levels are shifted dependent on their effective mass and the quantization length. In addition to this single-particle effect, the exciton binding energy is also modified. Here, we can discern two mechanisms: the reduced dimensionality leads to an increase of the exciton binding energy, as the degrees of freedom are reduced. For example, changing the dimensionality in the hydrogen problem from 3D to strict 2D confinement yields an increase of the binding energy by a factor of 4~\cite{Shinada-2DHydrogen}. In real low-dimensional structures, the change in dimensionality is often also accompanied by a change in dielectric environment. While in the bulk phase, there is an isotropic, typically rather large dielectric constant screening the attractive Coulomb interaction, it becomes anisotropic and smaller if the nano-structure is surrounded by vacuum. Together, these two effects lead to significant increases of the exciton binding energy and can stabilize excitons in nanostructures at elevated temperatures.

Apparently, in our picoperovskites, the increase of interband transition energy due to the 1D confinement and the increase of exciton binding energy nearly cancel out each other, as the shift only amounts to \qty{36}{\milli\electronvolt}. A similar lack of drastic changes to the optical band gap with changing dimensionality due to a compensation of these effects was previously reported for Boron Nitride nano-tubes~\cite{Wirtz-BNtubes}.

The PL map shown in \figref{fig:opto-pl}, obtained by moving the sample under the laser spot shows the source of the PL signal highly localized in a shape similar to the one seen in the optical image in a) and a homogeneous distribution of PL intensity. For the polarization patterns shown in d) Gaussian fits were performed for every angle, for the backscattered laser light reference at \qty{3.06}{\electronvolt} and for the PL at \qty{2.313}{\milli\electronvolt}. We clearly see that the PL emission of the picoperovskite bundle is not polarized parallel to the laser, but preferably oriented along its long axis. The emission pattern is again that of a two-fold symmetric system, but with an intensity decrease of roughly \qty{60}{\percent}, which is less pronounced than in the polarized Raman spectroscopy with a decrease of \qty{83}{\percent}. 

\begin{figure}
    \centering
    \includegraphics[width=\linewidth]{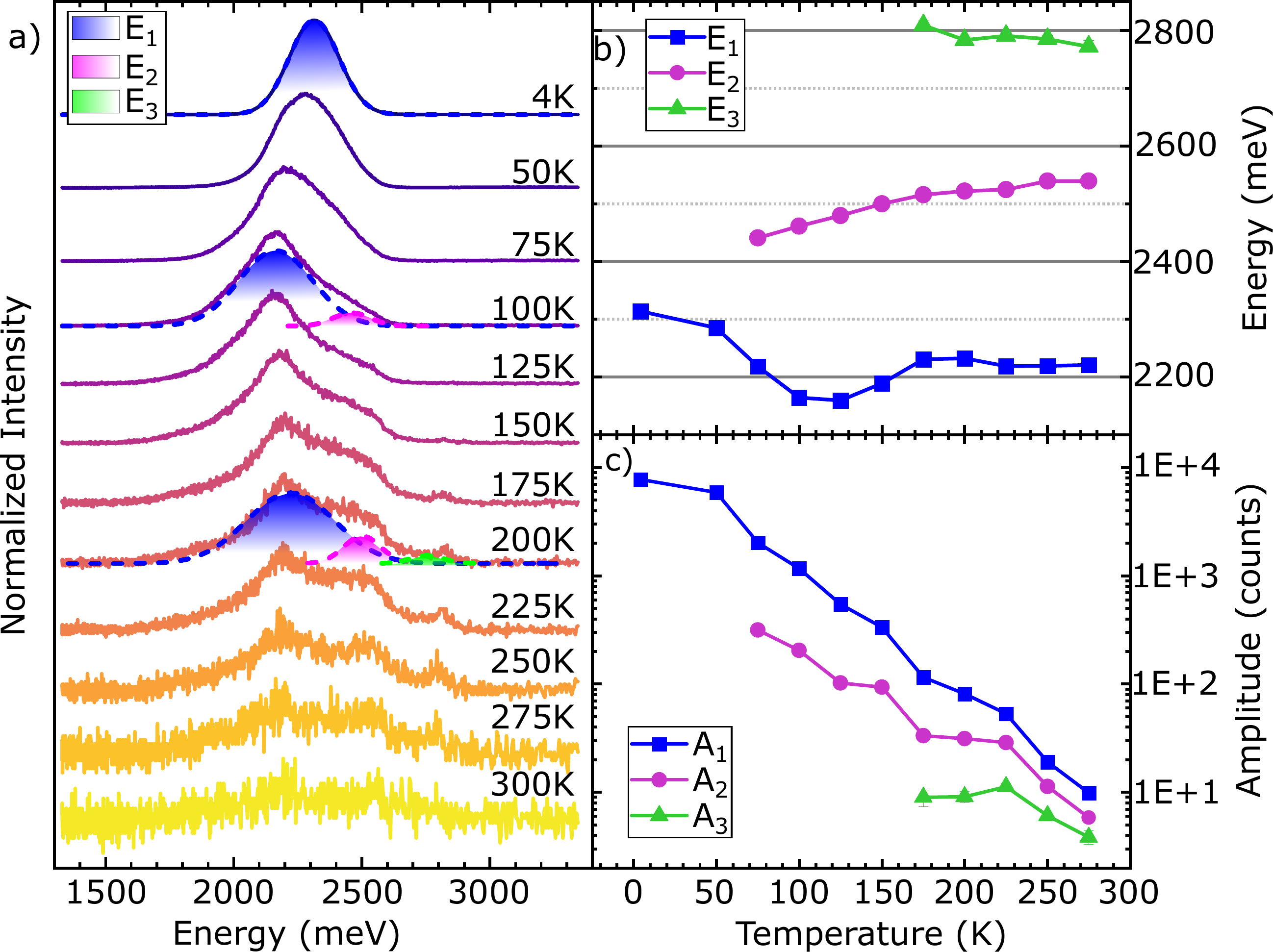}
    \caption{a) PL spectra taken at different temperatures with exemplary Gaussian fit shown for the spetra obtained at T=4, 100, 200 K. b) Peak energy of Gaussian fits performed on the spectra shown in a). The numbering of E$_{1,2,3}$ follows the same as for T=200 K in a). d) Amplitude of the Gaussian fits.}
    \label{fig:temp-pl}
\end{figure}

Due to the small dimension of the picoperovskite bundles, the PL signal is strongly position dependent (see SI~Figure~2). To perform a temperature series on the PL setup, influence of spatial inhomogeneities and temperature-induced drift of the sample must be overcome. To achieve this for each temperature, a PL map was obtained, and a single spectrum from approximately the same position was selected and fitted with Gaussian functions. \figref{fig:temp-pl} a) shows a waterfall plot of these curves, normalized, with three exemplary fits. For the temperature range of $T=\qty{4}{}-\qty{50}{\kelvin}$ one peak was identified, for $T=\qty{75}{}-\qty{150}{\kelvin}$ two and for \qty{175}{\kelvin} until \qty{300}{\kelvin} three peaks were identified. Note that fits could only be performed from \qty{4}{}-\qty{275}{\kelvin} due to the significantly lower signal-to-noise-ratio at room temperature. This trend becomes clear in \figref{fig:temp-pl} c), where the amplitudes of the Gaussian fits of all three peaks are plotted over temperature on a logarithmic intensity scale. The decrease in amplitude of over 3 orders of magnitude is consistent with the low intensity of fluorescence in \figref{fig:fluo-raman} a), which was measured at room temperature. 

\figref{fig:temp-pl} b) shows the energy of all three peaks, obtained from the Gaussian fits, plotted over the temperature. E$_2$ shows a linearly increasing blueshift of \qty{98,2}{\milli\electronvolt} between \qty{75}{} and \qty{300}{\kelvin}. Many perovskite materials show this unusual blueshift, which is not present in most other semiconducting materials. Generally speaking, there are two contributions to the change of gap energy with temperature. One is the change of electron-phonon interaction, e.g. by increasing the phonon population with rising temperature, leading to a decrease of the bandgap\cite{Lee2017, Zhang2020, Woo2018}.  The other is the thermal lattice expansion resulting in an increase of the bandgap. In contrast to conventional semiconductors, for ABX$_3$ perovskites the conduction band minimum and the valence band maximum are both formed by bonding BX and antibonding BX$^*$ orbitals\cite{Tang2021}. A change of bond length, -energy or orbital overlap, will lead to a change of bandgap. Although the blueshift is still under intense debate, it is apparent that the influence of thermal expansion has a significant contribution to it. The high energy peak E$_3$ shows a slight redshift between \qty{175}{} and \qty{275}{\kelvin} unusual for a perovskite material. As the peak only emerges with higher temperatures and a redshift is typically associated with an increase of electron phonon interaction, the origin in this high energetic peak might be higher band phonon-assisted transitions. Recent studies on perovskite nanocrystals have shown the opposite behavior for chlorine instead of iodine based perovskites\cite{Fasahat2025}, with an electron-phonon-coupling mediated redshift. The most unique behavior is shown by E$_1$ with a nonlinear redshift from 4 to \qty{125}{\kelvin} followed by a nonlinear blueshift until \qty{175}{\kelvin}, after which the energy doesn't change significantly. This peak dominates the spectrum at low temperatures and shows the most drastic decrease in intensity. A similar behaviour was reported for different MHPs thin films, like MaPbI$_3$ at low temperature which Wright et al attributed to phase transitions from orthorhombic to tetragonal between 130 and \qty{160}{\kelvin}\cite{Wright2016}, although the shift here is much larger. Due to the high intensity and large FWHM of \qty{230}{\milli\electronvolt} E$_1$ can be attributed to so called self trapped excitons (STEs)\cite{He2023}. Through the formation of excitons the lattice deformation is locally enhanced leading to self trapping of the excitons. As STEs usually have a long lifetime due to their high localization a temperature series on time resolved PL (TRPL) is carried out, to further solidify this claim.

\begin{figure}
    \centering
    \includegraphics[width=\linewidth]{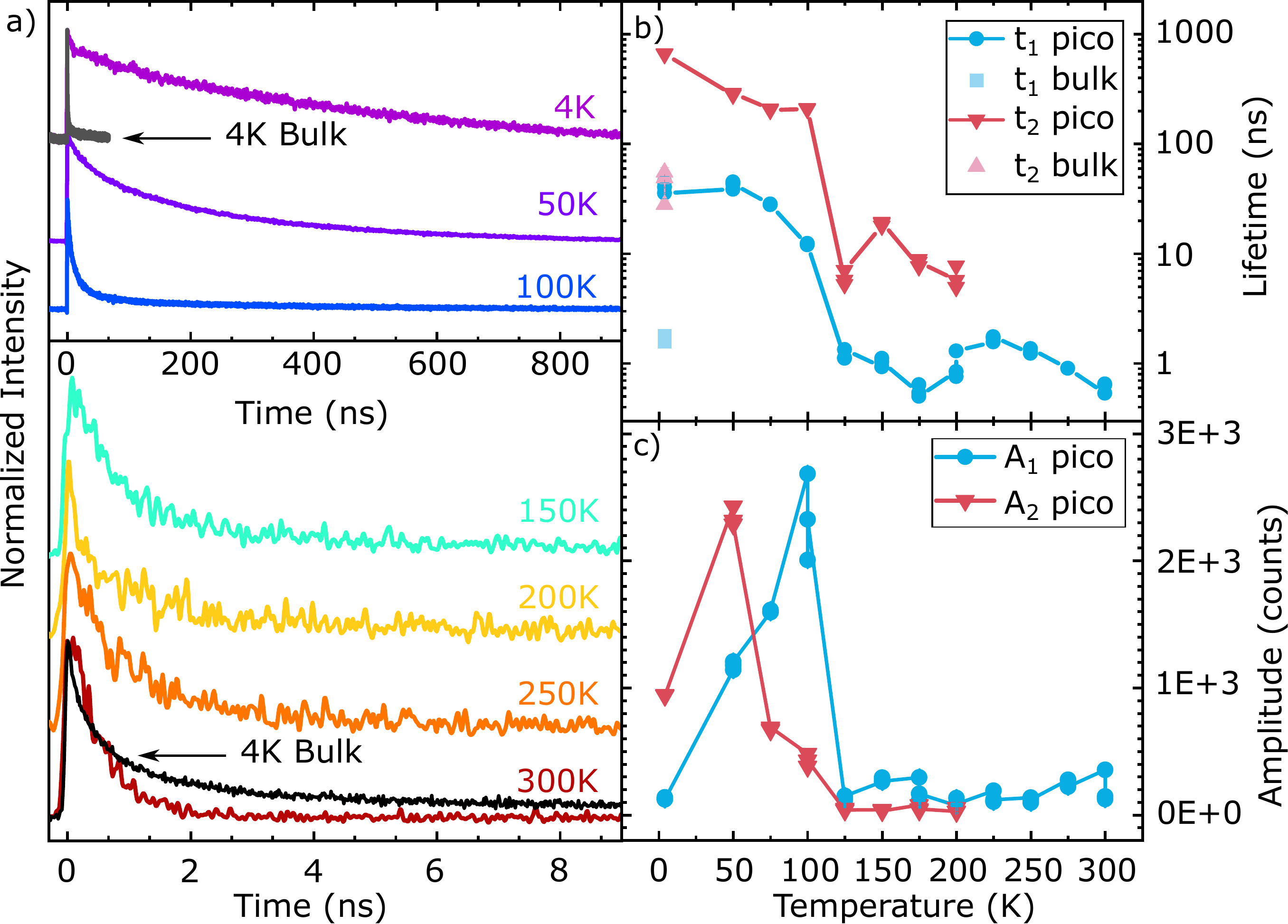}
    \caption{a) PL traces measured at different temperatures and a PL trace of a bulk sample at T=4 K for comparison. b,c) Lifetimes and Amplitudes obtained by biexponential fits performed on the traces from a).}
    \label{fig:temp-trpl}
\end{figure}

For time-resolved PL measurements, the same laser system as in the steady-state PL setup was used, but the diode laser was operated in pulsed mode, with an excitation wavelength of \qty{405}{\nano\meter} and a repetition rate of \qty{10}{\mega\hertz}. For detection a time-correlated single photon counting systems using an avalanche photodiode detector was coupled to the laser, experimental details are published elsewhere~\cite{Kempf-Perovskite}. The resulting time-dependent PL intensity traces can be seen in \figref{fig:temp-trpl} a), with two different timescale regimes, depending on the temperature. At low temperatures from 4 to \qty{100}{\kelvin}, the PL dynamics occurs on timescales of hundreds of nanoseconds, while at higher temperatures from 125 to \qty{300}{\kelvin} we observe shorter-lived signals on few-nanosecond timescales. 

The effect of changing dimensionality is especially pronounced in the PL curves measured at \qty{4}{\kelvin} (violet trace). Here, the picoperovskite sample has a PL lifetime exceeding that of the  bulk reference sample (black trace) by more than two orders of magnitude. Such long lifetimes have also been observed in other quasi-1D nanostructures, such as GaAs-based core-shell nanowires~\cite{Flo-Nanowires-spin} and associated, in part, with changes of spontaneous emission rate of excitons due coupling to the nanowire optical mode~\cite{Flo-Nanowires-tuning}. With its fully 1-dimensional character, the picoperovskite potentially represents the extreme limit of this mechanism. The fact that such long lifetimes can be observed experimentally is rather surprising, as PL lifetimes of excitons in SWCNTs have been observed to be far smaller, even at low temperatures~\cite{Marie-CNT-PL-Lifetime}. For CNTs, nonradiative processes provide additional decay channels and reduce the  PL lifetime to values far below the radiative lifetime while also reducing the quantum yield~\cite{Heinz-CNT-lifetime-PRL04}. 
Remarkably, the SWCNT  encapsulating the perovskite nanowire inside apparently suppresses such nonradiative decay channels for the perovskite inside. From the long PL lifetime we observe we can also infer that there is neither pronounced charge nor energy transfer from the perovskite core to the SWCNT shell, both of which would present a decay channel for the perovskite PL emission. Since we can assume based on CNT diameters the SWCNT band gap to be smaller than that of the perovskite, directly opposite to the usual design of core-shell structures, such transfer processes should be energetically favorable.

To systematically study the temperature dependence of the PL lifetimes, each curve was fitted with either a biexponential or a monoexponential decay function. For the biexponential function two contributions are attributed with drastically different timescales of decay. \figref{fig:temp-trpl} b) visualizes these two timescales on a logarithmic scale. The results from the short-lived timescale are denoted by the subscript 1 and plotted in blue. The resulting lifetimes are harshly contrasted to the results of the long-lived signal, denoted by the subscript 2 and plotted in red, with the difference exceeding one order of magnitude. In pale red and blue the lifetimes t$_1$ and t$_2$ of a bulk sample obtained at \qty{4}{\kelvin} are plotted for reference. Both values are decreased by one order of magnitude compared to the results from the picoperovskite sample. At \qty{4}{\kelvin} $t_2$ amounts to an astonishing \qty{659}{\nano\second}, decreasing to \qty{208}{\nano\second} at \qty{100}{\kelvin} followed by a significant drop of lifetime to \qty{7}{\nano\second} remaining in that order of magnitude until only a monoexponential fit can be carried out after \qty{200}{\kelvin}, e.g. because the decay channel for t$_2$ vanishes or a technical reason like the amplitude being to low to differentiate the signal. $t_1$ behaves similarly, starting with \qty{45}{\nano\second} at \qty{4}{\kelvin} dropping to roughly \qty{1}{\nano\second} after \qty{100}{\kelvin} which is approaching the temporal resolution threshold. The increase of $t_1$ after \qty{200}{\kelvin} can be attributed to the change from bi to monoexponential fitting, thus a technical reason. The huge PL lifetime and stark decrease with increasing temperature is a good indication of the presence of STEs. The decrease of PL lifetime with temperature might be associated with an increased phonon density or even an increased charge mobility, reducing the localization of the STEs. \figref{fig:temp-trpl} c) shows the amplitude derived from fitting, showing the relative share of each contribution $A_1$ and $A_2$. For 4 and \qty{50}{\kelvin} $A_2$, the longlived signal is dominant, which is overtaken by $A_1$ at \qty{75}{\kelvin}, the temperature for which in the steady state PL the second peak emerges (\figref{fig:temp-pl}). After \qty{100}{\kelvin} both amplitudes drop multiple orders of magnitude, the same temperature for which the drop in lifetime in b) occurs.

\section{DFT calculations}
\begin{figure}
    \centering
    \includegraphics[width=\linewidth]{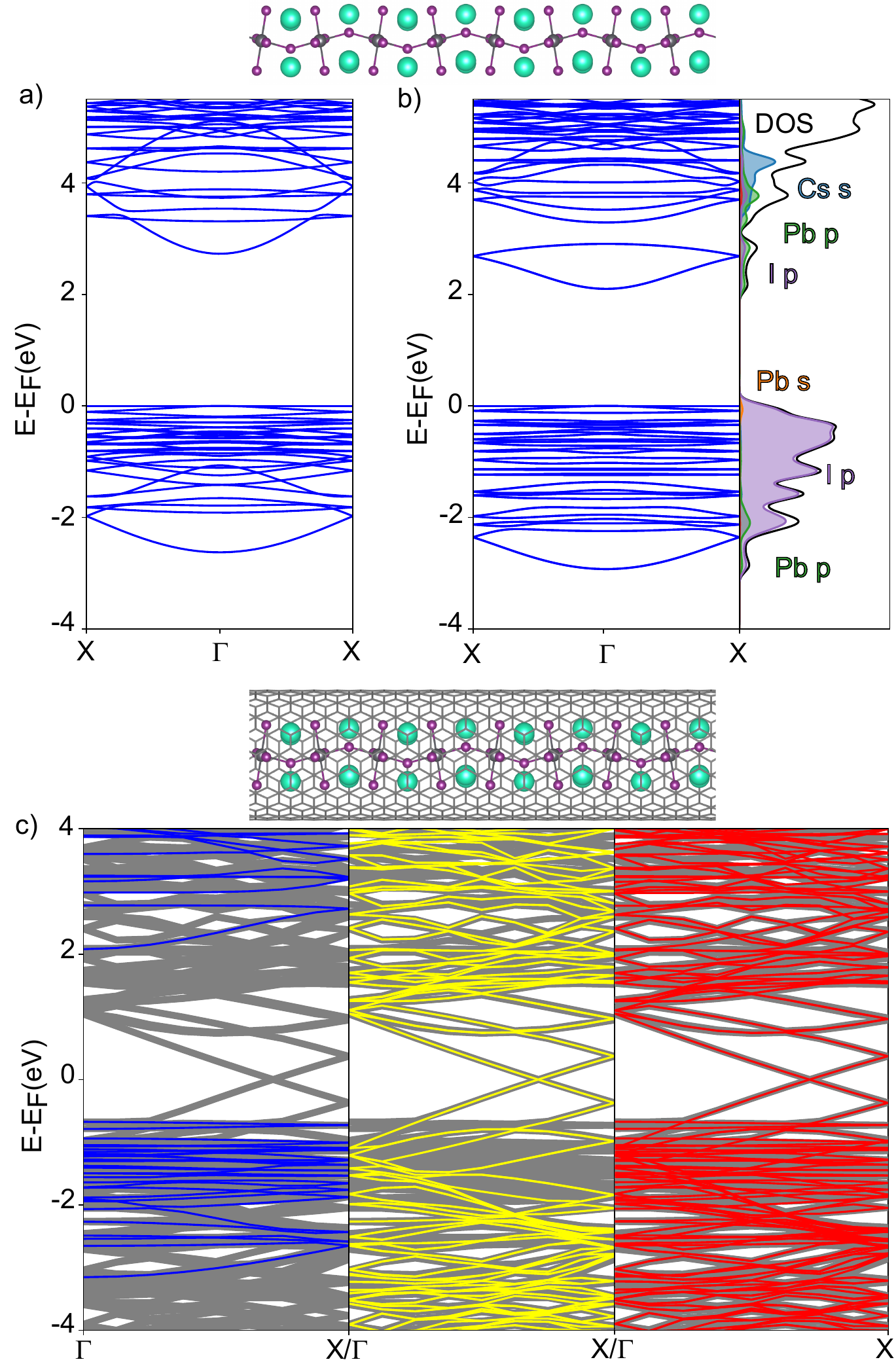}
    \caption{Electronic band structures calculated at the DFT-PBE level.  a) and b) show the band structures calculated for the isolated Cs$_3$PbI$_5$ picoperovskite, without a)  and with b) spin-orbit coupling (SOC).  b) also shows the bands' projections onto atomic orbitals (PDoS). c) shows the band structures of the isolated picoperovskite (blue), isolated CNT (yellow) and of the combined picoperovskite/CNT structure (red, and gray bands appearing in all panels).}
    \label{fig:dft}
\end{figure}

Density-functional theory calculations were carried out in
order to investigate the electronic structure of the combined
picoperovskite/CNT structure, where the picoperovskite is modelled as Cs$_3$PbI$_5$ which is charge neutral and resembles the cubic perovskite structure with Cs vacancies located at opposing cube corners. The band gap of the
Cs$_3$PbI$_5$ picoperovskite in vacuum is calculated
to be 2.73 eV, using the generalized-gradient PBE functional \cite{Perdew1996, Giannozzi2009}and
neglecting spin-orbit coupling (SOC).  Including SOC reduces
the gap to 2.09 eV, whilst using the more computationally-expensive
PBE0 hybrid functional \cite{Adamo1999} increases the gap by approximately 1 eV \cite{Kashtiban2023, Vona2022}.
The projected density-of-states (PDoS, \figref{fig:dft} b)) confirms that, 
as with higher-dimensional perovskite materials~\cite{Blancon2020}, the valence band is
primarily of I5p character, whilst the conduction band is a mix
of I5p and Pb6p contributions.  

Placing this picoperovskite inside  a (10,10) nanotube (diameter 1.36 nm) 
produces the band structure shown in \ref{fig:dft} c) (gray and red lines).  
The negligible hybridization between the picoperovskite and CNT
can be deduced by comparing this band structure to those calculated 
for the isolated picoperovskite, or isolated CNT 
[blue and yellow lines in \ref{fig:dft} c)].  We see that the bands of the
picoperovskite/CNT structure are essentially a superposition
of the bands of the isolated components, with each electronic 
state clearly identifiable as belonging to the picoperovskite
or the CNT.  This property allows the identification of the 
picoperovskite band edges in the combined structure, giving
a picoperovskite band gap of 2.74 eV, essentially identical
to the isolated case.

An interesting feature of the band structures in \figref{fig:dft} is the alignment of the band edges. Both the valence band maximum and conduction band minimum (VBM/CBM) of the picoperovskite lie energetically higher than the quasi VBM/CBM of the CNT(we note that the metallic (10,10) CNT
does have metallic bands which cross the Fermi level, but
their contribution to the DoS is small). This is comparable to the type-II band alignment of for example transition metal dichalcogenide based heterostructures \cite{Jiang2021}, in which excitons can be separated by the heterojunction creating interlayer excitons. These interlayer excitons have strongly increased PL lifetimes, compared to the intralayer excitons and also a pronounced temperature dependence. Like the PL lifetimes measured on the picoperovskites the PL lifetimes of interlayer excitons decreases with temperature, ranging from up to \qty{100}{\nano\second} at \qty{4}{\kelvin} to single nano seconds at room temperature\cite{Miller2017}. Since the CNTs are optically inactive in the used wavelength range, electrons are only excited in the picoperovskite site. If the heterojunction between CNT and picoperovskite allows for charge transfer from the perovskite to the CNTs an electron might get delocalized in the CNT while a highly localized hole resides in the picoperovskite site. This spatial separation would drastically reduce the wavefunction overlap and thus increase the lifetime of these excited states drastically.  However, the experimental system of this study is not as well defined as the model, with a range of CNTs of different diameters and chiral indices, over which the optical measurements are intrinsically averaged. The method of DFT calculations itself also has its limits, in that it does not capture excitonic effects.

\section{Summary and Conclusion}
In this paper optical studies on a new 1D perovskite material were presented, called picoperovskite because of its small diameter. This novel material was investigated in the form of a unique heterostructure, where the picoperovskite nanowires were grown inside of single wall carbon nanotubes. The investigated structures consisted of heterostructures of different lengths forming a bundle. The near-perfect encapsulation tackled one of the problems for performing optical experiments on nanoscaled perovskite materials and device application: degradation. The improved stability leads to a significantly reduced storage complexity and increased reproducability.

For the results of the optical experiments, we suggest two different explanations. At low temperatures, approaching \qty{4}{\kelvin} the steady-state PL is dominated by the formation of STEs, with their high FWHM of \qty{230}{\milli\electronvolt} and high intensity increase. Also the energy and charge transfer from the perovskite core to the CNT shell becomes insignificant, increasing the PL lifetime over two orders of magnitude and enhancing the self trapping mechanism. The compensation of quantization energy and increased exciton binding energies leads to the negligible PL shift between bulk and picoperovskites.

An alternative explanation is the emergence of charge transfer excitons, where one of the charge carriers is excited from the perovskite core to the CNT shell, analogous to the behaviour in TMD heterostructures. In these type-II heterostructures, the lifetime of interlayer excitons is significantly higher than that of the intralayer excitons, where the electron-hole pair resides in one of the materials. Analogously, an exciton in the CNT shell could recombine quickly, either radiatively or not, while single charge carriers could reside over a longer period. The spatial separation of the charge carriers would decrease the wavefunction overlap, thus increasing the PL lifetime. In this scenario the quantization energy would be partially compensated by the exciton-binding energy and the type-II band alignment. 

Based on the available data, neither model can be conclusively favored. However, since the investigated bundles consist of CNTs with varying chiral indices, and corresponding bandgaps only a subset of the bundles should be aligned according to the DFT calculations, also broadening the PL. With the methods and setups used here, it is not possible to verify whether only a subset of the bundles is emitting PL. To confirm one or the other explanation further research is necessary.

Finally, the unusual red- and blueshift of different peaks with temperature is of scientific interest and warrants further study, as it might be originating from novel interactions between excitons and phonons interplaying with crystal phase changes.

\section{Methods}
\textbf{Material synthesis and sample fabrication}
CsPbI$_3$ was prepared by solid-state synthesis using mixtures of CsI and PbI$_2$. All precursors were sourced from Merck in their maximum available purity (99.999\%) and stoichiometric mixture was finely ground using an agate mortar and pestle inside an argon-atmosphere glove box. The mixture was transferred into 10 cm silica quartz ampoules that were sealed under vacuum ($10^{-3}$ Torr) using an oxygen-methane torch. The sealed ampoules were transferred into a muffle furnace, heated at 2 °C min$^{-1}$ to 480 °C and held for 5 h and cooled to room temperature at 1 °C min$^{-1}$. The as prepared CsPbI$_3$ was deposited inside SWCNTs, supplied by NanoIntegris, with a specified diameter range of 1.2-1.7 nm according to a published melt protocol\cite{Sloan1999}. 10 mg of SWCNTs was oxidized in open air at 480 °C before being intimately mixed in a mortar and pestle with separate \qty{30}{\milli\gram} CsPbI$_3$ before being placed in 10 cm ampoules sealed under vacuum ($10^{-3}$ Torr). The CsPbI$_3$ and SWCNT mixture was then heated to 525 °C for a 12 h dwell period employing ramp rates of 2 °C min$^{-1}$ and cooling rates of 1 °C min$^{-1}$. The sample was ultrasonically dispersed in ethanol and drop-cast onto 3.05 mm lacey carbon-coated copper grids (Agar) for examination by electron microscopy.

A double-corrected JEM-ARM 200F microscope operating at 80 kV equipped with CEOS imaging aberration and probe correction for Annular dark-field scanning transmission electron
microscopy (ADF-STEM) investigations was used for high-resolution analysis of the heterostructures after synthetization. ADF-STEM images were obtained with a JEOL annular field detector with a fine imaging probe with a convergence semi-angle of a ~25 mrad, a probe current of \qty{23}{\pico\ampere} and an inner angle of 45-50 mrad.

The picoperovskite filled SWCNTs were shipped in a vacuum sealed glass flask, and later stored under ambient conditions. The source powder (filled CNTS/Bulk perovskite) was thinned out in an isopropanol solution and sonicated for several minutes. This helped loosen up big bundles of material. This colloidal solution was then dropped onto a cleaned Si/SiO$_2$ substrate and treated in a spin coater for a few minutes. To identify fitting accumulations the substrate with the thin film was then investigated using a fluorescence microscope to analyze the material distribution. The fabrication was considered successful if at least one spot could be identified with a good contrast of fluorescence, isolation of the bundle and reasonably straight orientation.

\textbf{Steady state and time resolved optical characterization}
For the steady-state PL measurements, the sample was excited with a \qty{3.06}{\electronvolt} (\qty{405}{\nano\meter}) continuous-wave diode laser using a power of \qty{10}{\micro\watt}, focused to a spot size of about \qty{1}{\micro\meter} with a 50x long working distance objective. The sample was cooled to a nominal temperature of \qty{6}{\kelvin}. The actual temperature of the sample can be higher. The PL light emitted by the sample was collected using the same objective, filtered through a \qty{450}{\nano\meter} (\qty{2.755}{\electronvolt}) long-pass filter, and analyzed with a combination of a spectrometer and a charge-coupled device. To obtain the PL map of the sample (\figref{fig:opto-pl}), the cryostat, with the sample inside, was moved in relation to the fixed laser spot through a computer-controlled xy stage. The polarization-resolved measurements were achieved by placing a linear polarizer and an achromatic half-wave plate ($\lambda/$2) in the excitation and detection beam path.

The time-resolved PL was measured with the same laser setup, but a slightly higher excitation power at \qty{30}{\micro\watt} and a repetition rate of \qty{10}{\mega\hertz}. The temporal pulse width of the laser used in that power regime was around \qty{50}{\pico\second}, so the instrument response function was mostly defined by the detection, yielding a time resolution of around \qty{500}{\pico\second} (see the instrument response function in \cite{Kempf-Perovskite}. Here, the fundamental laser wavelength was filtered out by a \qty{450}{\nano\meter} (\qty{2.755}{\electronvolt}) long-pass filter, while the PL signal was measured through a commercial single photon avalanche diode using time-correlated single photon counting electronics. 

For the steady-state Raman measurements, the sample was excited with a \qty{532}{\nano\meter} continuous-wave diode-pumped solid-state laser using a power of \qty{2}{\milli\watt} focused to a spot size of about \qty{1}{\micro\meter} with a 100x long working distance objective. The Raman scattered light was collected by the same objective and filtered by 3 reflective volume Bragg gratings (BragGrate(tm) by Optigrate) with a central wavelength of \qty{532}{\nano\meter} and analyzed with a spectrometer. For the polarization-resolved measurements, a linear polarizer and an achromatic half-wave plate ($\lambda/$2) were placed in the detection line.

\textbf{Cryogenic measurements}
All optical measurements were performed in a flow cryostat with a temperature variability from 4 to \qty{300}{\kelvin}. Inherent to this cooling technique, the sample was kept under a vacuum for the whole measurement time and thus protected from environmental degradation by oxygen and water. Between measurements, samples were kept in ambient conditions.

\textbf{DFT calculation}
Density-functional theory calculations were carried out using  plane wave basis sets and ultrasoft pseudopotentials, as  implemented in the Quantum ESPRESSO software package \cite{Giannozzi2009, DalCorso2014}. For the isolated picoperovskite, a simulation cell consisting of two Cs$_3$PbI$_5$ units was constructed, with the two Cs vacancies located  at their most stable positions, i.e.\ at opposite corners, as found previously for Cs-Pb-Br \cite{Kashtiban2023}.  The details of the computation, such as cutoff energies, reciprocal space sampling, vacuum regions and  force and pressure thresholds, are the same as described in Ref. \cite{Kashtiban2023}. The combined picoperovskite\ CNT structure was constructed from five repeats  of a (10,10) CNT and two Cs$_3$PbI$_5$ units.  The simulation cell was fixed to the CNT value and the Cs$_3$PbI$_5$ units were relaxed within this cell.  We note that building the cell in this way led to a relatively small strain on the picoperovskite, with the CNT cell having length \qty{12.325}{\angstrom} and the ideal vacuum cell of the picoperovskite having length \qty{12.253}{\angstrom}.  After the relaxation, the band structure of the combined structure, and also of the isolated CNT and picoperovskite, were calculated.  Note that the picoperovskite was not re-relaxed after the CNT was removed, which leads to small differences in the band structures shown in \figref{fig:dft} c).

\section{Author Information}
\textbf{Corresponding Authors}\\
\textbf{Reza Kashtiban} - Department of Physics, University of Warwick, Coventry, CV4 7AL UK;\\
Email: r.kashtiban@warwick.ac.uk\\
\textbf{Tobias Korn} - Institute of Physics, University of Rostock, Rostock, 18059, Germany\\
Email: tobias.korn@uni-rostock.de\\

\textbf{Authors}\\
\textbf{Maximilian Tomoscheit} - Institute of Solid State Physics, Friedrich Schiller University Jena, 07743 Jena, Germany\\
Email: maximilian.tomoscheit@uni-jena.de\\
\textbf{Julian Schröer} - Institute of Physics, University of Rostock, 18059 Rostock, Germany\\
\textbf{Jaskaran Singh Virdee} - Institute of Physics, University of Rostock, 18059 Rostock, Germany\\
\textbf{Rico Schwartz} - Institute of Physics, University of Rostock, 18059 Rostock, Germany\\

\section{Acknowledgements}
M.T. gratefully acknowledges fruitful discussion with G. Soavi. M.T., J.S. R.J.K are indebted to the EPSRC(UK) for support from grant IAA 2022-2026(1029) and M.T. and T.K.  acknowledge financial support by the DFG \emph{via} SFB1477 (project No. 441234705). CEP is grateful for computational support from the UK national high performance computing service, ARCHER2, for which access was obtained via the UKCP consortium and funded by EPSRC grant ref EP/X035891/1.

\section{Keywords}
1D perovskites, optical characterization, photostability, exciton dynamic, carbon naotubes

\section{conflict of Interest}
The authors declare no conflict of interest.

\section{Data Availability Statement}
The data that support the findings of this study are available from the corresponding author upon reasonable request.

\FloatBarrier
\newpage
\bibliographystyle{ieeetr}
\bibliography{perovskites}
\clearpage
\newpage

\section{Supplementary information}

\begin{figure*}
	\centering
	\includegraphics[width=\textwidth]{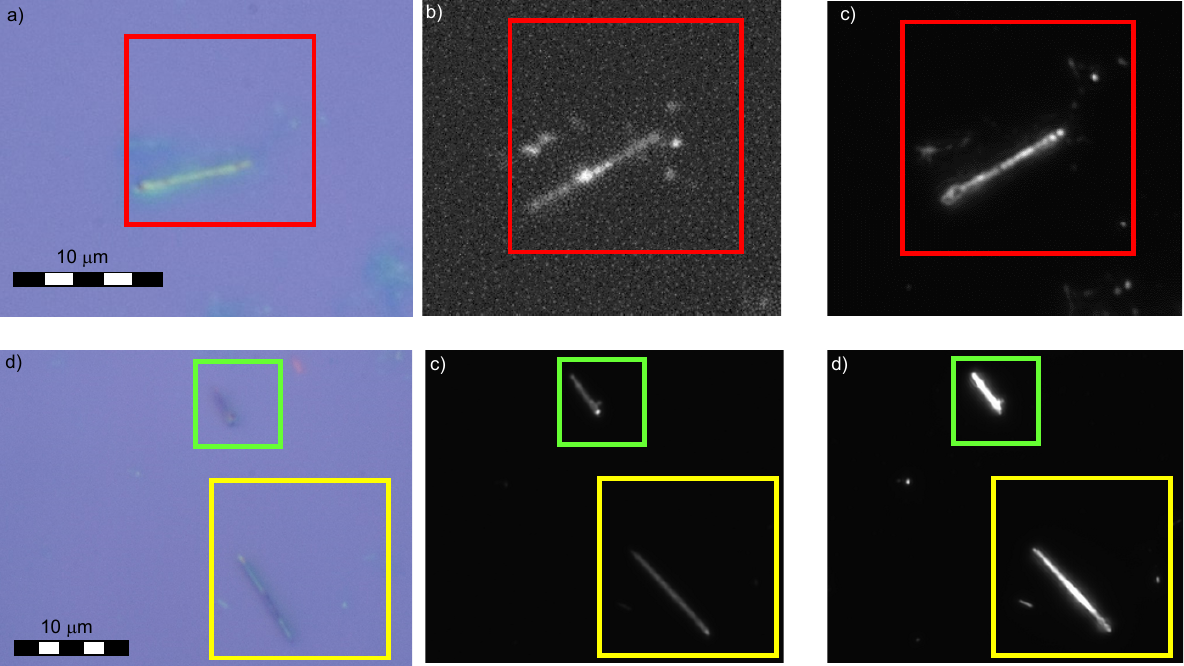}
	\caption{\textbf{S1.} a), b), c) Optical, Fluorescence, Dark field image of sample 1. d), e), f) Optical, Fluorescence, Dark field image of sample 2 and 3.}
	\label{sifig:images}
\end{figure*}

\begin{figure*}
	\centering
	\includegraphics[width=\textwidth]{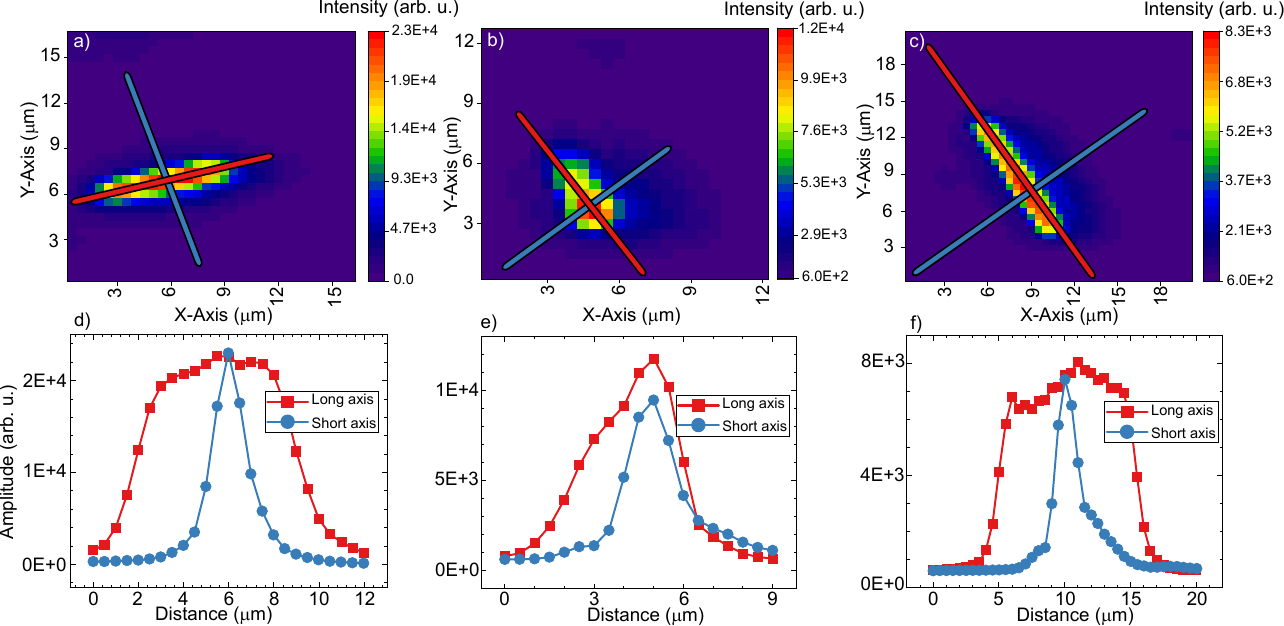}
	\caption{\textbf{S2.} a), b), c) False color map of PL intensity extracted from a scan of the area marked by the boxes in \textbf{Figure \ref{sifig:images}} on sample 1, 2 and 3. d), e), f) Linescans obtained from the maps shown in a), b) and c) noted by the red and blue colored bars.}
	\label{sifig:pl}
\end{figure*}

\begin{figure*}
	\centering
	\includegraphics[width=\textwidth]{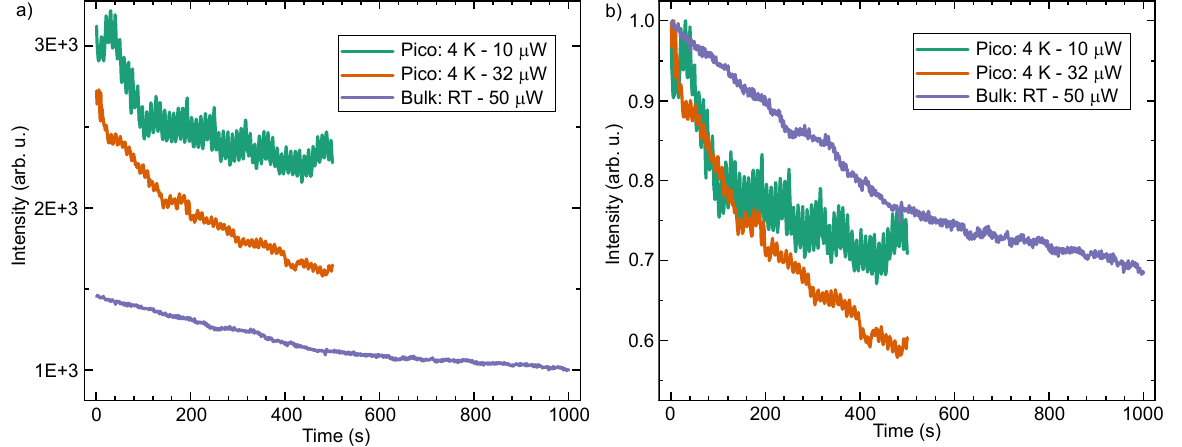}
	\caption{\textbf{S3.} Amplitudes obtained by fitting multiple PL spectra with 1 s integration time under constant laser exposition performed on sample 1 for two different laser powers and a bulk sample for comparison. a) absolute, b) normalized values.}
	\label{sifig:degrad}
\end{figure*}

\begin{figure*}
	\centering
	\includegraphics[width=\textwidth]{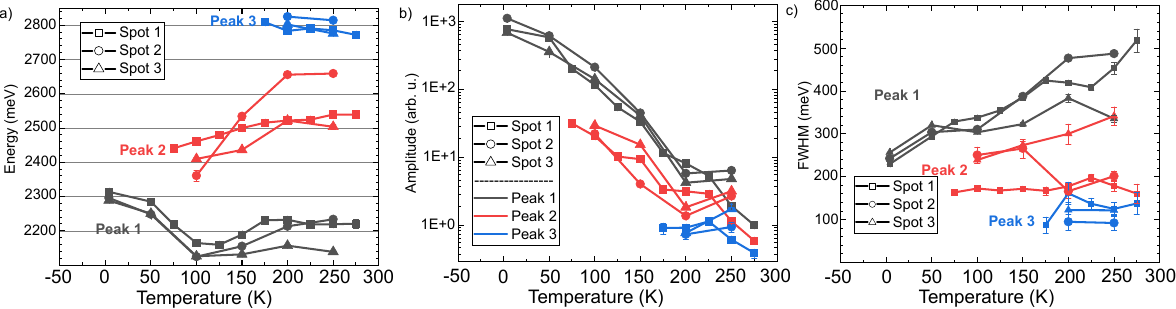}
	\caption{\textbf{S4.} a) Energy, b) Amplitude and c) FWHM of a PL temperature series for all 3 samples.}
	\label{sifig:pl-temp}
\end{figure*}

\begin{figure*}
	\centering
	\includegraphics[width=\textwidth]{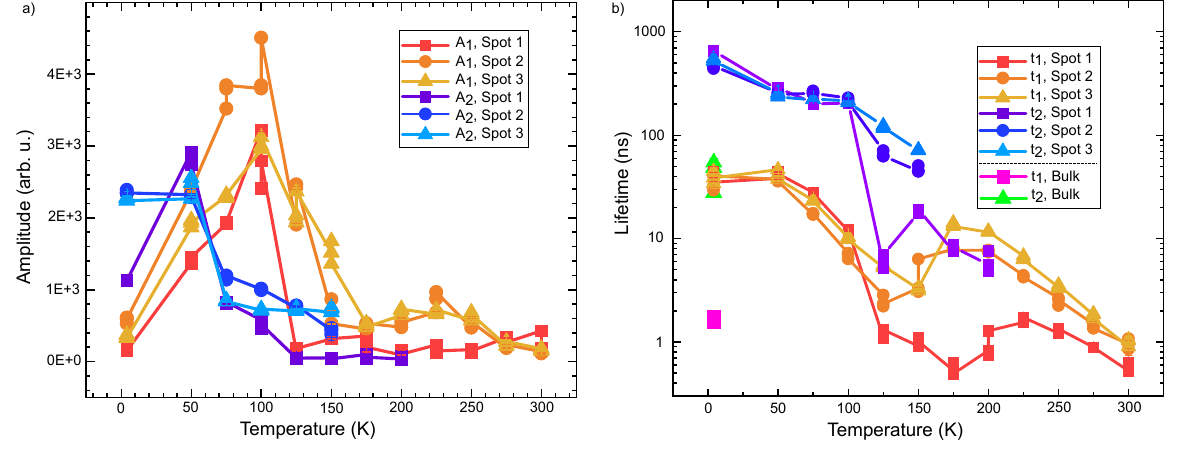}
	\caption{\textbf{S5.} a) Amplitude and b) Lifetime from TRPL temperature series for all 3 different samples and the results from 1 Bulk sample at 4K.}
	\label{sifig:trpl-temp}
\end{figure*}
\end{document}